\documentclass[twocolumn,showpacs,showkeys,preprintnumbers]{revtex4}
\usepackage{graphicx}
\usepackage{subfigure}
\usepackage{lscape}
\usepackage{dcolumn}
\usepackage{bm}
\usepackage{color}

\usepackage{amsmath} 
\usepackage{amssymb} 
\usepackage{multirow}
\usepackage{epstopdf}

\begin{document}
\title{Sagittarius $\mathrm{A}^{\star}$ as a plausible source candidate for PeV neutrinos}
\author{Sabyasachi Ray}
\email{methesabyasachi@gmail.com}
\author{Rajat K. Dey}
\email{rkdey2007phy@rediffmail.com}
\affiliation{Department of Physics, University of North Bengal, Siliguri, WB 734 013 India}

\begin{abstract}
We propose that the recently observed diffuse neutrinos by IceCube with energies above 1 PeV might have originated from Sagittarius $\mathrm{A}^{\star}$ located in the galactic disk. This implies that the astrophysical settings of Sagittarius~$\mathrm{A}^{\star}$ need to accelerate hadronic cosmic rays to energies of $\sim 100$ PeV or more. Then, the hadronic emission scenario argues that this galactic neutrino source is also a PeV gamma-ray source. Recent observation of galactic diffuse PeV gamma-rays with energies $\sim 1$~PeV by Large High Altitude Air Shower Observatory has also advocated this conjecture. In the present paper, we demonstrate that if protons are accelerated to energies of $\sim 100$~PeV or more as reported by Osmanov {\it et al.} (Astrophys. J. {\bf 835} 164:2017) in Sagittarius $\mathrm{A}^{\star}$ environment, then they might generate PeV neutrinos and gamma rays through cosmic rays-gas/interstellar matter ({\it e.g.} $pp$) interactions. We estimate theoretically the diffuse neutrino flux due to back-to-back charged pion decays and the accompanying gamma-ray flux from neutral pion decays. These results suggest that a fraction ($\simeq 1\%$) of the PeV diffuse neutrino flux observed by IceCube can be explained by the neutrino emission from Sagittarius $\mathrm{A}^{\star}$. Upcoming IceCube Gen2 and Cherenkov telescope array could be able to test our scenario for PeV neutrino and gamma-ray emissions from the only known galactic Pevatron Sagittarius $\mathrm{A}^{\star}$ with CR energies more than $100$~PeV.  

\end{abstract}

\keywords{Galactic center, particle acceleration, plasma motion, neutrinos, gamma rays}
\maketitle

\section{Introduction}	
The origin of cosmic rays (CRs) and the associated particle acceleration mechanisms are among the most theoretically challenging issues in astroparticle physics. Usually, the principal component protons ($p$) of the energetic CRs will collide with ambient matter and radiation via $pp$ and $p\gamma$ hadronic interactions within their source regions [1-2]. Consequently, high-energy astrophysical neutrinos and gamma rays ($\gamma$-rays) are produced [1-2]. The end products as neutrinos and $\gamma$-rays may typically carry $\sim 3-5\%$ and $\sim 8-10\%$ energy of the accelerated CR proton, respectively [3]. This indicates that the observed PeV scale neutrinos in IceCube [4] hints the existence of some hitherto unknown astrophysical settings where CRs could be accelerated around $\sim 100$~PeV scale. This energy scale falls very much within the galactic-to-extragalactic transition region of the CR primary energy spectrum [5].

It is believed that galactic CRs only contribute a small amount beyond the CR knee to the CR energy spectrum. The generated neutrinos and $\gamma$-rays from potential $pp$- and $p\gamma$- interactions may have energies ranging from a few hundred TeV to even considerably higher [6]. However, detecting PeV scale neutrinos by IceCube boosts the community to search for galactic and extragalactic sources of CRs with energies around $100$ PeV and beyond. Sources like microquasars [7], shell-type supernova remnants (SNRs) [8], pulsar wind nebulae [9], Galactic Halo [10], Fermi Bubbles [11], or Sagittarius $\mathrm{A}^{\star}$ [12] (Sgr $\mathrm{A}^{\star}$) were considered to be promising galactic candidates of TeV - PeV scale neutrinos [13].  

Deep $\gamma$-ray observations surrounding the galactic center (GC) by the High Energy Stereoscopic System (H.E.S.S.) observatory [14] detected $\gamma$-rays close to $\sim 35$~TeV energy scale in the bounds of Sgr $\mathrm{A}^{\star}$. These diffuse $\gamma$-rays could have possibly been produced via interactions of galactic PeV CRs originating from a Pevatron Sgr A$^{\star}$ with a dense gaseous zone encircling the Sgr $\mathrm{A}^{\star}$, instead of the more often theorized SNRs [15]. Detection of TeV neutrinos is a must for the validation of the hadronic origin of the diffuse H.E.S.S. TeV energy $\gamma$-rays from the center of Sgr $\mathrm{A}^{\star}$. The galactic center, along with the Sgr $\mathrm{A}^{\star}$ as plausible neutrino sources, were considered previously in several works [16-17].

A major challenge, however, is to detect ultra-high-energy (UHE) $\gamma$-rays that are produced outside of the Milky Way owing to strong attenuation by cosmic microwave background radiation (CMBR) and infrared (IR)  background. On the other hand, many sources emitting some few tens of TeV to few PeV $\gamma$-rays, as reported by the recent detections [18-19], have opened up the possibility of Pevatrons with CR energies in the range 1 - 100 PeV residing in our galaxy. The recent observed spectral energy distribution (SED) of $\gamma$-rays just above 10 TeV and up to 1 PeV from the galactic source, LHAASO J1908+0621 can only be explained with the hadronic component provided by the source [18-19].
 
It is a well-known fact that a high-energy $\gamma$-ray source could be an accelerator of hadronic CRs. In confirming these high-energy CR accelerators, the detection of PeV energy neutrinos would be a smoking gun. Over a decade of search for some galactic PeV neutrino sources, however, no direct evidence has been found [20-22]. In this paper, we present a viable possibility that some PeV energy neutrinos observed by IceCube may have come from Sgr $\mathrm{A}^{\star}$, the supermassive black hole (SMBH) positioned at the dynamical center of our galaxy. Protons having much energy ($\sim 100$~PeV or more) generated by some acceleration mechanism in Sgr $\mathrm{A}^{\star}$ environment may interact with nearby protons/gas or radiative photons or interstellar medium (ISM) [23]. However, the ambient photon density in Sgr $\mathrm{A}^{\star}$ is relatively low, but the target proton density is very high. In addition, other features of Sgr $\mathrm{A}^{\star}$, such as x-ray and near-infrared (NIR) flares and neutrino bursts, indicate a starburst-like environment in the galactic plane [24]. Consequently, the time scale of $pp$ interaction is tiny compared to the $p\gamma$ interaction for PeV neutrino production [25]. In our present study, we calculate the PeV energy neutrino and $\gamma$-ray fluxes from Sgr $\mathrm{A}^{\star}$ resulting from $pp$ interactions only. In principle, we assume that the source environment of Sgr $\mathrm{A}^{\star}$ is suitable for the working of the Langmuir-Landau-centrifugal Drive (LLCD) acceleration mechanism [23]. Then, Sgr $\mathrm{A}^{\star}$ can accelerate protons to $\sim 100$~PeV or more, and the PeV neutrino production is very likely. These neutrinos are expected to be detected by the upcoming detectors [26-27]. Non-detection of galactic neutrino sources to date might be due to poor sensitivities of detectors to the dimmer TeV-PeV neutrino emission.  

In [23], Osmanov {\it et al.,} calculated that protons from the Sgr $\mathrm{A}^{\star}$ might reach up to energy $170$~PeV. Protons with this much energy will interact with background photons and the ambient gas. Also, the presence of the distribution of molecular hydrogens in the galaxy makes the $pp$ channel more efficient than the other one. These $pp$ interactions will generate an almost equal number of pion species ($\pi^0,\pi^+,\pi^-$). $\gamma$-rays result from the decay of $\pi^0$, while the decay of charged pion pairs produce three neutrinos, three anti-neutrinos and electrons together. In the mechanisms, $p\rightarrow \pi^0 \rightarrow 2\gamma$ and $p\rightarrow \pi^\pm \rightarrow 3\nu+3\tilde{\nu}$, the number ratio is $\gamma:\nu+\tilde{\nu} \sim 1:3$ and the corresponding energy ratio is $\frac{E_{\gamma}}{E_{\nu}} \sim 2$ [3]. Hence for the protons of $E_{p}\sim 100$~PeV from Sgr $\mathrm{A}^{\star}$, the maximum possible $\gamma$-ray and $\nu$ energies from these process will be, $E_{\gamma} \sim \frac{1}{10}E_{p}=10$ PeV and $E_{\nu} \sim \frac{1}{20}E_{p}=5$ PeV respectively.

There should be two essential conditions that Sgr $\mathrm{A}^{\star}$ and its environment must simultaneously satisfy. First, the source Sgr $\mathrm{A}^{\star}$ must accelerate protons/hadrons to $\sim 100$~PeV or more. Secondly, the Sgr $\mathrm{A}^{\star}$ should have an ambient column density with appropriate thickness or a gas number density (for neutrino point source) or diffuse ISM number density (for neutrino diffuse source). The two-step LLCD mechanism is a likely scenario that may lead to manoeuvre protons at energies $\sim 100$~PeV or higher [23]. These protons are expected to be dispersed isotropically from the source via the very efficient LLCD acceleration mechanism. Therefore, a reasonably high gas density as targets in the direction of $\sim 100$~PeV proton beam is desired.

Suppose CRs with energy $\geq 100$~PeV (although the total energy released from a single outburst is $\sim 3\times 10^{54}$~erg [28-29]) were injected by the powerful outburst of Sgr $\mathrm{A}^{\star}$ more than $10^{6}-10^{7}$~years ago. In that case, they might have distributed in the galactic halo. A fraction of these CRs might have also escaped from the galaxy during their diffusion in the halo. For a homogeneous spherical halo containing the production region of $\nu/\gamma$-rays, $R_{\text{eff}}$  accounts for the effective halo radius centred at Earth [30]. The {\lq All Sky\rq} model [31] picked $R_{\text{eff}}=10$~kpc corresponding to the target density $n_{\text{H}}\approx 1.6$~cm$^{-3}$ for the best fit of the calculated neutrino spectra to the data. Here, $n_{\text{H}}$ is the average hydrogen number density of galactic matter. On the other hand, the H.E.S.S. field encompassing some dense molecular regions around the galactic center gave evidence of hydrogen molecules with $n_{\text{H}} \sim 10$~cm$^{-3}$ and $n_{\text{H}} \sim 120$~cm$^{-3}$ as obtained from data [32-33].

The energetic protons generated by the LLCD mechanism participate in the hadronic interactions with the distributed $\text{H}_2$ (mainly) gases surrounding Sgr $\mathrm{A}^{\star}$ and/or ISM. These interactions at the end contribute galactic PeV neutrinos and $\gamma$-rays. The study only takes into account the issue of the possibility of PeV neutrino and $\gamma$-ray emissions from $\mathrm{A}^{\star}$ and the theoretical estimation of the corresponding neutrino/$\gamma$-ray fluxes. As the $\text{H}_2$ gas number density can be as high as 120~cm$^{-3}$ in the region surrounding $\mathrm{A}^{\star}$, it is therefore much conceivable that the $pp$ processes are populated over the $p\gamma$ processes. It should be noted that the emission of neutrinos and $/$or $\gamma$-rays from pp interactions in the neighbourhood regions of Sgr $\mathrm{A}^{\star}$ also come from the galactic plane. 

The outline of the paper is as follows. The key features of the LLCD mechanism are briefly highlighted in the next section. The third section contains a compact calculation of the PeV neutrino flux and the accompanying PeV $\gamma$-ray flux. The last section summarizes the conclusions. 
	
\section{Acceleration of protons via LLCD mechanism in Sgr $\mathrm{A}^{\star}$}
Some previous works realized that the current rate of particle acceleration in and around Sgr $\mathrm{A}^{\star}$ is insufficient to accelerate protons to PeV energies or more [34]. However, it is possible to attain particle energies up to $\sim 100$~PeV range, if the proton acceleration rate in Sgr $\mathrm{A}^{\star}$ could be of the order of $\geq 10^{38}$~erg~s$^{-1}$ [35-36]. Rapid variability, mostly in x-rays and sometimes in $\gamma$-rays, was seen during flares in Sgr $\mathrm{A}^{\star}$ [37]. Moreover, the past jet activity of Sgr $\mathrm{A}^{\star}$ can form the Fermi bubbles below and above the galactic plane [38]. Authors in [39],  advocated that some of the IceCube neutrino events might have origins in the Fermi bubbles. In Sgr $\mathrm{A}^{\star}$s' flaring epoch, the luminosity may even exceed $\sim 100$-fold compared to its current state, where it stays very often. IceCube neutrino events, either as preceding or following the powerful flares in Sgr $\mathrm{A}^{\star}$, are likely to be correlated [40]. Hence, we propose the $\mathrm{A}^{\star}$ as a plausible supplier of energetic protons with energies $\sim 100$~PeV or more. All the very often exploited particle acceleration scenarios [41-43] did not facilitate any CR productions to such high energies. As an alternative particle acceleration scenario, the LLCD mechanism, based on a series of some linear and nonlinear processes in an electron-proton plasma medium around the SMBH at the galactic center region, {\it i.e.} the Sgr $\mathrm{A}^{\star}$ could accelerate protons to the maximum energy $\sim 170$~PeV [23] and references therein. Therefore, we will draw attention to several key aspects of the LLCD mechanism that make it easier for protons to achieve the desired energy. 

Protons can be boosted up to $\sim 100$~PeV energies or more in a standard two-step LLCD particle acceleration mechanism. The transformation of SMBH's gravitational energy to the final particle energy is unfolded via the two-step workings of the LLCD in the magnetized plasma surrounding the SMBH at Sgr $\mathrm{A}^{\star}$ [23].        

An idealized description of magnetized plasma would be as two streams of protons and electrons. In the first step of the LLCD, the free energy inherent in the varying behaviour of plasma particles to the SMBH's gravity intensifies a rapidly growing linear instability in Langmuir waves at a rate, 

\begin{equation}
 \Gamma_{\text{GR}}=\frac{\sqrt{3}}{2}\left(\frac{\omega_e{\omega_p}^2}{2} \right)^{\frac{1}{3}} J_{\mu}(b)^{\frac{2}{3}} 
\end{equation}

where $J_\mu$, being the Bessel's function of the first kind, and $\mu=(\frac{\omega_{e}}{\Omega})$. Also, $\omega_{e,p}=\sqrt{4\pi{e^2}{n}_{e,p}/{m}_{e,p}{\gamma_{e,p}^3}}$ is the plasma frequency in which $\gamma_{e,p}$ accounts the Lorentz factor associated with electrons and protons. For the two species, parameters $m_{e,p}$ and $n_{e,p}$ denote their rest masses and number densities, respectively. Here, $\Omega$ is the SMBH's angular velocity of rotation, which was approximated to $\frac{ac^3}{GM}\approx 2.5\times{10^{-3}}\frac{a}{0.05}$~rad~s$^{-1}$ [44-45]. The rotation rate is characterized by the parameter $a$, which has a range $0-1$. In Eq. (1), the argument of $J_\mu$ is $b=\frac{2ck}{\Omega}\sin{\phi_{ep}}$ with $\phi_{e,p}={(\phi_{p}-\phi_{e})/2}$, the half-value of the initial phase difference between the species $p$ and $e$, and $k$ is the usual wave vector. The centrifugal force that drives the electrostatic waves in the $e-p$ plasma is unequal for the two species, so the condition, $\gamma_{e}\neq \gamma_{p}$ is being sustained.

The local magnetic field strength where $e-p$ plasma co-rotates with the SMBH during the active phase of Sgr $\mathrm{A}^{\star}$ in the past is [46]

\begin{equation}
B\approx \sqrt{\frac{2L_{\text{BL}}}{r^2c}}\approx 15.4\times{(\frac{L_{\text{BL}}}{5\times{10^{38}}{\text{erg}~\text{s}^{-1}}})}^{\frac{1}{2}}\times\frac{10R_{\text{S}}}{r}~\text{G}, 
\end{equation} 

where $R_{\text{S}}$ denotes the Schwarzschild radius of the SMBH, and $L_{\text{BL}}$ is the bolometric luminosity (BL) of Sgr $\mathrm{A}^{\star}$ during its past active phase. For a magnetized plasma in Sgr $\mathrm{A}^{\star}$'s environment, the Larmor radii of plasma species should be several orders of magnitude smaller than $R_{\text{S}}$. This situation allows plasma particles to move along the field lines. The plasma species suffer extremely efficient acceleration with linear velocity of rotation, $v\rightarrow c$,  in the region close to the light cylinder zone $r\approx R_{\text{lc}}$ ($R_{\text{lc}}=\frac{c}{\Omega}$ is the light cylinder radius).

The magnetic field continuously regulates the centrifugally driven energetic particles to move along straight field lines until the plasma energy density overtakes the magnetic field. The energy in the amplified waves (Langmuir waves) thus reaches an optimum level which is quite below the PeV energy protons reported in [47]. However via the final step of the LLCD mechanism these group of energetic protons can generate a group of highly energetic protons in the $e-p$ plasma in a spherically symmetric accretion disk at the galactic center. As a result, a new bunch of protons will get out in the region, upon which the Landau damping of the Langmuir waves will transpire, and eventually protons entering into the PeV energy scale. We consider a specific case for which $\gamma_{p}\sim 10^{3}$, the corresponding instability time-scale $\sim \frac{1}{\Gamma_{\text{GR}}}$ is much smaller than the kinematic time-scale $\sim \frac{1}{\Omega}$ or even the cooling time-scale (possible energy loss processes by protons). This also ensures the equality between the instability growth rate ($\Gamma_{\text{GR}}$) for the generation of Langmuir waves and the rate of Landau damping ($\Gamma_{\text{LD}}$). All these indicate that the pumping of Langmuir waves is a very efficient process in transforming gravitational energy into particle kinetic energy. 

Through the final step, Landau damping (LD) caused by a constructive Langmuir collapse of Langmuir waves, the local beam of protons receives enormous energy. At this stage, the phase velocities of these waves overtake the speed of light if $\phi_{ep}>\frac{\pi}{6}$. Then, no protons are left in the magnetosphere to unceasing Landau damping. Some previous calculations identified the region in the magnetosphere with $r < R_{\text{lc}}$, where the Langmuir waves do not encounter any collapse [48]. On the contrary, the effects of the magnetic field in the region with $r > R_{\text{lc}}$ becomes insignificant for the Langmuir collapse. Nevertheless, the accretion processes are vital in the region ($r > R_{\text{lc}}$) to retain the particle density. In [49], the number density of protons in the case of spherically symmetric accretion is given by,

\begin{equation}
n = \frac{L_{\text{BL}}}{4\eta_{\text{c}}\pi{m_{p}}c^2vR_{\text{lc}}^2}\approx 6.3\times 10^{4}~\text{cm}^{-3}~.
\end{equation} 

Here, $10\%$ of the rest mass energy of the accreting matter was expected to contribute to emission. In Eq.(3), the $L_{\text{BL}}\approx 5\times 10^{38}$~erg~s$^{-1}$  and $\eta_{\text{c}}\simeq 0.1$ are used. Besides, the high-frequency pressure $\rm P_{\text{f}}\propto |E|^{2}$ varies as $\frac{1}{l}$ manner inside the magnetosphere where the magnetic field is so strong to push protons along the lines of the force [48]. On the contrary, the thermal pressure in the region, obeying a $P_{\text{th}}\propto\frac{1}{l^{2}}$ variation, increases much faster than $P_{\text{f}}$. This situation does not favour the eventuality of the Langmuir collapse inside the magnetosphere, and these waves progress towards the exterior part of the magnetosphere. As the magnetic field of the exterior part does not constrain the particles' motion, then the $P_{\text{f}}\propto \frac{1}{l^3}$ variation is maintained [48]. The faster rise of $P_{\text{f}}$ over the $P_{\text{th}}$, ensuring the Langmuir collapse to begin [50-51].

The final stage Langmuir collapse, boosts the electrostatic field, thereby transferring the field energy to the local beam of protons. The protons gain enormous energy, resulting in the final step of the LLCD process, it is given by [51],  

\begin{equation}
\epsilon_{p} \approx \frac{ne^{2}}{4\pi^{2}\lambda_{\text{D}}^{3}}\Delta{r}^{5}~(\text{eV}).
\end{equation}

where $\Delta{r}$ is a finite radial distance scale corresponding to the radial behaviour of the Lorentz factors $\gamma_{r}$ of accelerated protons [52], and $\lambda_{\text{D}}=\sqrt{K_{\text{B}}T_{\text{e}}/4{\pi}n_{0}e^{2}}$ is the Debye length scale with $K_{\text{B}}=1.38\times 10^{-16}$~erg~K$^{-1}$, and $T_{\text{e}}=10^{10}$~$K$ (Sgr $\mathrm{A}^{\star}$ is in radiatively inefficient accretion flow (RIAF) state in the outburst phase [52]), being the electron temperature. In addition, the Goldreich-Julien number density is, $n_{0}=\frac{\Omega{B}}{2\pi{ec}}$ with $B\approx 15.4$~G, and $R_{\text{lc}}\approx 1.19\times{10}^{13}$~cm. We use $\Delta{r}\approx \frac{R_{\text{lc}}}{2\gamma_{p}}$ [49]. Substituting all the relevant parameters and some of their values into Eq.(4), and after some rearrangements of terms in it, we get       

\begin{equation}
\epsilon_{p} \approx 35\times{\frac{\eta_{\text{c}}}{0.1}}\times{(\frac{a}{0.05})^{1/2}}\times{(\frac{T_{\text{e}}}{10^{10}~\text{K}})^{-3/2}}\times{(\frac{10^{3}}{\gamma_{p}})^{5}}\times{({L_{\text{BL},39}})^{5/2}}, 
\end{equation}
in PeV.\\   
It can be shown using Eq.~(5) that protons could achieve energies close to $\approx 198$~PeV for a possible arbitrary set of parameters; $\eta_{\text{c}}=0.1$, $a=0.05$, $\gamma_{p}=500$ and $L_{\text{BL}}= 5\times 10^{38}$~erg~s$^{-1}$ (here, $L_{\text{BL},39}\equiv \frac{L_{\text{BL}}}{(10^{39}~\text{erg}~\text{s}^{-1})}$). It is noteworthy to mention that such a smallest value of $\gamma_{p}=500$ is enough to satisfy the essential conditions, $\Gamma_{\text{GR}}\approx \Gamma_{\text{LD}}$ and $\frac{1}{\Gamma_{\text{GR}}}<<\frac{1}{\Omega}$ for the efficient working of the LLCD mechanism. Hence, each $\nu$ and $\gamma$ from $pp$ interaction could receive the maximum possible energies $\sim 9.9$ and $\sim 19.8$~PeV respectively [3]. 

\section{PeV neutrino and accompanying $\gamma$-ray fluxes} 

The $pp$ interactions produce $\pi^{\pm}$ and $\pi^{0}$ mesons, and their subsequent decays result in the production of PeV neutrinos and $\gamma$-rays. In the active phase of Sgr $\mathrm{A}^{\star}$, protons must disperse arbitrarily via the LLCD acceleration mechanism in the region. In case of such an isotropic flux of energetic protons from Sgr $\mathrm{A}^{\star}$, the emissivity (pion numbers~vol$^{-1}$~sr$^{-1}$) of $\pi^{0}$/$\pi^{\pm}$ can be presented by an analogous  equation of the form mentioned below [53-54],

\begin{equation}\label{key}
	Q_{\pi^0}^{pp}(E_{\pi^0})=cn_{\text{H}}
	\int_{E_p^{\text{th}}}^{E_p^{\text{max}}}\frac{dn(E_p)}{dE_p} \frac{d\sigma_{pp}(E_{\pi^0},E_p)}{dE_{\pi^0}} dE_p
\end{equation}
where, $n_{\text{H}}$ is the ambient ISM/hydrogen number density. The inelastic $pp$ interaction cross-section is borrowed from [54] with the projectile energy dependence as $\sigma_{pp}(E_{p})=34.3 +1.88~ln[E_p/1~\text{TeV}] +0.25[ln(E_p/1~\text{TeV})]^2$~mb. This parametrization was found with the numerical data fitting in the code by the authors of the high energy hadronic interaction model, {\it Sybill} [55]. In Eq.~(6), $\frac{d \sigma_{pp}(E_{\pi^0},E_p)}{dE_{\pi^0}}$ accounts the differential cross-section for an inelastic fixed-target $pp$ collision for the production of $\pi^{0}$ with energy $E_{\pi^0}$. Exploiting a parametrization for the inelastic differential cross-section according to the so-called scaling model [56], the conversion relation appears as     
  
\begin{equation}
	\frac{d \sigma_{pp}(E_{\pi^0},E_p)}{dE_{\pi^0}} \simeq\frac{\sigma_{pp}}{E_{\pi^0}}{f_{\pi^0 }(x)}
\end{equation}
where, $x\equiv E_{\pi^0}/E_p$. The function ${f_{\pi^0}(x)}$ takes the standard expansion in $x$ followed from the work [3].  

Exploiting Eq.~(7) and the power-law for the CR differential flux: $\frac{dn(E_p)}{dE_p}\propto E_{p}^{-\alpha}\equiv \frac{dn(E_{\pi}^{0})}{dE_{\pi}^{0}}\propto E_{{\pi}^{0}}^{-\alpha}$ ($\alpha$ being the spectral index), the $\pi^o$ emissivity can be expressed as

\begin{equation}
	Q_{\pi^0}^{pp}(E_{\pi^0})= cn_{\text{H}}\sigma_{pp}\frac{dn(E_p)}{dE_p} \times Z_{p\pi^0}(\alpha)
\end{equation}

where $Z_{p\pi^0}(\alpha)$ is the spectrum-weighted  moment of the inclusive cross-section, which is given by [3],
\begin{equation}
	Z_{p\pi^0}(\alpha)=\int_{0}^{1}x^{\alpha-2}{f_{\pi^0}(x)}dx
\end{equation}
\\

The $\gamma$-ray emissivity results from $\pi^{0}$-decays, produced in $pp$ interactions is then found in terms of $\pi^{0}$ emissivity by the following
  
\begin{equation}
	Q_{\gamma}^{pp}(E_{\gamma})=2\int_{E_{\pi^{0}}^{\text{min}}(E_{\gamma})}^{E_{\pi^{0}}^{\text{max}}}\frac{Q_{\pi^{0}}^{pp}(E_{\pi^{0}})}{(E_{\pi^{0}}^{2}-m_{\pi^{0}}^{2})^{1/2}}dE_{\pi^{0}}
\end{equation}
 A further substitution with $E_{\pi^{0}}^{\text{min}}(E_{\gamma})=E_{\gamma}+\frac{m_{\pi^{0}}^2}{4E_{\gamma}}$ simplifies the above Eq.~ into   
\begin{equation}
	Q_{\gamma}^{pp}(E_{\gamma})\simeq Z_{\pi^o\gamma}(\alpha) Q_{\pi^0}^{pp}(E_{\pi^0})	
\end{equation}

where $Z_{\pi^0\gamma}(\alpha)=2/\alpha$, let us now denote it by $Z_0$.
\\
We know that the $pp$ process generates both the $e$-type and $\mu$-type of neutrino/anti-neutrino flavors ($e$-type: 2 numbers and $\mu$-type: 4 numbers). Just analogous to Eq.~(11), the corresponding neutrino and anti-neutrino emissivities for these above-mentioned flavours are given by

\begin{equation}
	Q_{\nu_{\mu}}^{pp}(E_{\nu_{\mu}})\simeq 2[Z_{\pi\nu_{\mu}}(\alpha)+Z_{\mu\nu_{\mu}}(\alpha)] Q_{\pi}^{pp}(E_{\pi})	
\end{equation}
\begin{equation}
	Q_{\nu_{e}}^{pp}(E_{\nu_{e}})\simeq 2Z_{\mu\nu_{e}}(\alpha) Q_{\pi}^{pp}(E_{\pi})	
\end{equation}

where factor 2 refers to the contribution from both the neutrino and anti-neutrino species. Here, we assume that the emissivities or energy distributions of $\pi^{0}$, $\pi^{+}$ and $\pi^{-}$  are identical with a given $\alpha$ {\it i.e.} $Q_{\pi^0}^{pp}(E_{\pi^0})\approx Q_{\pi^{\pm}}^{pp}(E_{\pi^{\pm}})$. 

Let us designate $Z_{\pi\nu_{\mu}}(\alpha)$ as $Z_1$, $Z_{\mu\nu_{\mu}}(\alpha)$ as $Z_2$ and $Z_{\mu\nu_{e}}(\alpha)$ as $Z_3$. These are the Z-factors accounting $\pi\rightarrow \nu_{\mu}$-decay, $\mu\rightarrow \nu_{\mu}$-decay and $\mu\rightarrow \nu_{e}$-decay respectively. These factors are evaluated through their forms in [54]. 

From the Eq.~(8), the total neutrino emissivity combining all the neutrino flavours into account is given by,

\begin{equation}
	Q_{\nu}=c{n_{\text{H}}}Z\sigma_{pp}\frac{dn(E_p)}{dE_p} \times Z_{p\pi^0}(\alpha)
\end{equation}
with $Z=2[Z_1+Z_2+Z_3]$\\

Hence, Eq.~(11) and Eq.~(14) give rise the $\gamma$-ray and neutrino emissivities from the $pp$ interaction process. 

The solution of CR transport equation indicates a uniform CR energy density ($N_{\text{CR}}$ in cm$^{-3}$~GeV$^{-1}$) spread over the entire galaxy, including the region surrounding the Sgr $\mathrm{A}^{\star}$. Thus the observed CR flux ($\phi_{\text{CR}}\approx \phi_{p}$) at energies close to $100$~PeV can be used to compute $N_{\text{CR}}\approx N_{p}=\frac{4\pi}{c}\phi_{p}=\frac{dn(E_p)}{dE_p}$, and also the expected $\gamma$-ray and neutrino fluxes. The expected $\gamma$-ray and neutrino fluxes would be 

\begin{equation}
\phi_{{\gamma}/{\nu}}=\frac{1}{4\pi} \int_{0}^{1}\frac{Q_{{\gamma}/{\nu}}}{4\pi{r}^2}dV=\frac{R_{\text{eff}}}{4\pi}\times{Q_{{\gamma}/{\nu}}}
\end{equation}
  
where $R_{\text{eff}}\equiv\int\frac{dV}{(4\pi{r}^2)}$ is the effective halo radius centred at Earth, which is calculated to be between $1$~kpc to $10$ kpc depending on the shape of the considered halo [30]. Finally, combining Eq.~(14) and Eq.~(15), the computed $\gamma$-ray and neutrino fluxes in terms of known parameters are written in the form 

\begin{equation}
	\phi_{\gamma}=n_{\text{H}}\sigma_{pp}Z_0 R_{\text{eff}}\phi_p \times Z_{p\pi^0}(\alpha)
\end{equation}
\begin{equation}
	\phi_{\nu}=n_{\text{H}}\sigma_{pp}Z R_{\text{eff}}\phi_p \times Z_{p\pi^0}(\alpha)
\end{equation}	

\subsection{Numerical results}

The mass composition of CRs at energies $100 - 300$ PeV from radio observations found a dominating light-mass nuclei (H and He), which might have a galactic origin [57]. Debate continues whether those CRs are linked with giant flares/outbursts of Sgr $\mathrm{A}^{\star}$. The energy-dependent diffusion coefficient of CRs in the galactic halo ascertains that these $100 - 300$ PeV CRs have already run away from the galactic halo when the most probable knee CRs get detected on Earth after $\sim 10^6$~years. This means the same outburst of Sgr $\mathrm{A}^{\star}$ about $\sim 10^6$~years ago could not provide the local CR flux level around and above the knee. A viable alternative would be some later outburst of Sgr $\mathrm{A}^{\star}$ from where CRs were injected with energies $\geq 100$~PeV.
 \begin{table*}
	\begin{center}
		\begin{tabular}
			{|l|l|l|r|} \hline
			$n_{\text{H}}$~{(cm$^{-3}$)}   & $\phi_{\nu}$~{(GeV$^{-1}$cm$^{-2}$s$^{-1}$sr$^{-1}$)}    & ${E_\nu}^2\phi_{\nu}$~{(GeV~cm$^{-2}$s$^{-1}$sr$^{-1}$)}    & ${E_{\gamma}}^2\phi_{\gamma}$~{(GeV~cm$^{-2}$s$^{-1}$sr$^{-1}$)}      \\    
			&&& \\ \hline
			
			$1$   & $1.4\times 10^{-27}$   & $1.0\times 10^{-13}$   & $4.6\times 10^{-13}$      \\ 
			&&& \\ \hline
			
			$10$   & $1.4\times 10^{-26}$   & $1.0\times 10^{-12}$  & $4.6\times 10^{-12}$      \\ 
			&&& \\ \hline
			
			$120$   & $1.6\times 10^{-25}$   & $1.1\times 10^{-11}$  & $5.5\times 10^{-11}$      \\ 
			&&& \\ \hline
		\end{tabular}
		\caption {Computation of the PeV neutrino and $\gamma$-ray fluxes for different values of $n_{\text{H}}$ with $R_{\text{eff}}\approx 10$~kpc, and ${\alpha}=3$.} 
	\end{center}
\end{table*}
SNRs are commonly believed to be the supplier of CRs with proton domination up to $\sim 3$~PeV energy [58-59]. The observed CR spectrum follows a power law with a spectral index $-3.0$ in the energy range $\sim 3 - 300$~PeV. Based on a standard diffusion-halo model [60-62], the CRs of this range may be attributed to a series of outbursts at Sgr $\mathrm{A}^{\star}$. The model predicted flux was lower than the observed CR flux data if the model parameters were tuned suitably in the calculation [62]. Their predicted CR flux $\phi_{p}$ dominated by protons, was close to $\sim 4\times 10^{-23}$~GeV$^{-1}$cm$^{-2}$s$^{-1}$sr$^{-1}$ at $\sim 100$~PeV [61]. Thus, the only Pevatron located in the galactic plane, and perhaps linked to the Sgr $\mathrm{A}^{\star}$, seems insufficient to explain the total CRs reaching Earth.

For $n_{\text{H}}$, we employ three possible choices described in the introduction [3]. The effective spherical-shaped halo radius can have a value up to $10$ kpc, and hydrogen density can be as high as $n_{\text{H}}= 120$~cm$^{-3}$. We also set $\sigma_{pp}=93.2038$~mb and $Z_{p\pi^0}(3.0)=0.021$ [54]. Thus, the corresponding maximum possible neutrino flux would be $1.1\times 10^{-11}$~GeV cm$^{-2}$s$^{-1}$sr$^{-1}$, which is included in Table 1. The accompanying $\gamma$-ray flux is found to be $5.5\times {10^{-11}}$~GeV~cm$^{-2}$s$^{-1}$sr$^{-1}$.

\section{Conclusions}

First, we have argued that the Sgr $\mathrm{A}^{\star}$ is a potential acceleration site of protons/nuclei. Protons could reach maximum energy, $\sim 198$~PeV, via two successive steps outlined in section 2. Secondly, this work has focused on the aftermath era of the acceleration process where these ultra-relativistic protons interact with an ambient gas or/and ISM, producing PeV neutrinos and $\gamma$-rays via the $pp$ hadronic channel. We summarize the main conclusions of the work as follows.

The acceleration rate of protons in Sgr $\mathrm{A}^{\star}$ could be of the order of $\sim 10^{38}$~erg~s$^{-1}$ in the past, and that would be sufficient to supply protons with energies $\sim 100$~PeV or more. The present study considers the possible roles of Sgr $\mathrm{A}^{\star}$ as a probable galactic Pevatron in CR production in the considered energy regime. Since the diffusion-halo model prediction on the CR flux level around the knee and up to $\sim 100$ PeV energies from Sgr $\mathrm{A}^{\star}$ is slightly lower compared to the local CR flux [5,61], the well-established SNR paradigm of galactic CRs is still being considered at least up to the CR knee. 

The computed diffuse PeV neutrino flux from Sgr $\mathrm{A}^{\star}$ here agrees well with some previous calculations on classified AGN, such as flat spectrum radio quasars (FRSQs) and BL Lacs [63-65]. Some strong constraints on our present estimations arise from $n_{\text{H}}$ and $R_{\text{eff}}$ in connection with ambient gas or ISM number density and the spherical-shaped halo radius. 

Simultaneous detection of PeV fluxes of neutrinos and $\gamma$-rays from the galactic center region close to  Sgr $\mathrm{A}^{\star}$ only can give evidence in favour of the generation of hadronic CRs with energies $\sim 100$~PeV or more. Our computed flux limits in this work are about two orders of magnitude lower than the present experimental limits based on the search using seven years of neutrino track data in IceCube [66]. This reiterates that only a tiny fraction ($\simeq 1\%$) of the observed diffuse astrophysical neutrinos may have their birth linked with Sgr $\mathrm{A}^{\star}$. A simultaneous operation of the upcoming Icecube-Gen2 [67] and the Cerenkov Telescope Array (CTA) [68] observatories would be capable of discriminating such a small fraction of PeV neutrinos of galactic origin from extragalactic ones. These near-future neutrino and gamma-ray observatories are expected to scan the galactic disk with better sensitivity and improved angular resolution. 


\end{document}